\documentclass[prc,a4paper,twocolumn,superscriptaddress,showkeys,showpacs]{revtex4}

\input{epsf}

\newcommand{\delslash}{\partial \hspace{-6pt}/}

\newcommand{\Nmassmed}{m_{N}^*} % mass of nucleon in medium
\newcommand{\Nmassvac}{m_{N}} % mass of nucleon in vacuum
\newcommand{\Nstarmassmed}{m_{N^*}^*} % mass of N* in medium
\newcommand{\Nstarmassvac}{m_{N^*}} % mass of N* in vacuum
\newcommand{\Nstarwidmed}{\Gamma_{N^*}} % width of N* in medium
\newcommand{\pimass}{m_\pi}  % mass of pion
\newcommand{\etamass}{m_\eta}  % mass of eta
\newcommand{\getaNR}{g_\eta} % eta NN* coupling
\newcommand{\gpiNR}{g_{\pi NN^*}^*} % pi NN* coupling
\newcommand{\gpiNRsq}{g_{\pi NN^*}^{*2}} % squared pi NN* coupling
\newcommand{\mevsig}{\langle \sigma \rangle} % nuclear matter
\begin{document}

\title{Medium effects to $N(1535)$ resonance and $\eta$ mesic
nuclei}
\date{\today}

\author{D. Jido}
\email{jido@rcnp.osaka-u.ac.jp}
\thanks{present address: 
        Research Center for Nuclear Physics (RCNP), 
        Ibaraki, Osaka 567-0047, Japan}
\affiliation{Departamento de F\'{\i}sica Te\'orica and IFIC, 
Centro Mixto Universidad de Valencia-CSIC,  
Ap. Correos 22085, E-46071 Valencia, Spain}
\author{H. Nagahiro}
\affiliation{ Department of Physics, Nara Women's University, Nara
630-8506, Japan}
\author{S. Hirenzaki}
\affiliation{Departamento de F\'{\i}sica Te\'orica and IFIC, 
Centro Mixto Universidad de Valencia-CSIC,
Ap. Correos 22085, E-46071 Valencia, Spain}
\affiliation{ Department of Physics, Nara Women's University, Nara
630-8506, Japan}

\begin{abstract}
The structure of $\eta$-nucleus bound systems ($\eta$ mesic
nuclei) is investigated as one of the tools to study in-medium properties of the
$N(1535)$ ($N^{*}$) resonance by using the chiral doublet model to
incorporate the medium effects of the $N^{*}$ resonance in a chiral
symmetric way.  We find that the shape and the depth of the
$\eta$-nucleus optical potential are strongly affected by the
in-medium properties of the $N^{*}$ and the nucleon.  Especially, as a general
feature of the potential, the existence of a repulsive core of the
$\eta$-nucleus potential at the nuclear center with an attractive part at the
nuclear surface is concluded.  
We calculate the level structure of bound states in this
'central-repulsive and surface-attractive' optical potential and find
that the level structure is sensitive to the in-medium properties of the
$N^*$.  The (d,$^3$He) spectra are also evaluated for the formation 
of these bound states to investigate the experimental feasibility.
We also make comments on the possible existence of
halo-like $\eta$ states in $\beta$-unstable halo nuclei.
 \end{abstract}

\pacs{12.39.Fe,14.20.Gk,14.40.Aq,25.10.+s,36.10.Gv}
\keywords{$\eta$ mesic nucleus, $N(1535)$, chiral symmetry, medium effect}

\maketitle

\section{Introduction}

The study of the in-medium properties of hadrons has attracted continuous
attention and is one of the most interesting topics of
nuclear physics.  
So far, several kinds of hadron-nucleus bound systems have been
investigated such as pionic atoms, kaonic atoms, and $\bar{p}$ atoms
\cite{batty97}.  The structure of these systems is described by means of a
complex optical potential, which reproduces well the experimental
data obtained by the X-ray spectroscopic methods
\cite{laat91}.  The search of the deeply bound pionic atoms has been
performed from the end of 80's
\cite{toki88,nieves90,iwasaki91,matsuoka95}, and the use of the recoilless
(d,$^3$He) reaction \cite{yamazaki96} led to the successful
discovery of the deeply bound states, which enable us to deduce in-medium pion
properties \cite{waas97,yamazaki98,itahashi00}.  Now extensions of
the method to other meson-nucleus bound states are being widely
considered both theoretically and experimentally \cite{tsushima98,
hayano99, klingl99, hirenzaki01}.

In this paper we consider the $\eta$ mesic nucleus as one of the doorways to 
investigate the in-medium properties of the $N(1535)$ ($N^*$).
The special features of the $\eta$ mesic nucleus are the following; 
(1) the $\eta$-$N$ system dominantly couples to $N(1535)$ at the
threshold region \cite{nstar}.
(2) The isoscalar particle $\eta$ filters out contaminations of the
isospin 3/2 excitations in the nuclear medium.
(3) Due to the $s$-wave nature of the $\eta NN^*$ coupling there is no
threshold suppression like the $p$-wave coupling.  
The strong coupling of the $N^*$ to $\eta N$ makes the use of this channel 
particularly suited to investigate this resonance in a cleaner way than the use of 
$\pi N$ for the study of other resonances like the $N(1440)$ and $N(1520)$.

%This $N^*$-dominance in $\eta$-$N$ system gives advantages in
%investigations of $N^*$ compared with those of other nucleon
%resonances through the $\pi$-$N$ system, for instance $N(1440)$ and
%$N(1520)$.

On the other hand, the in-medium properties of hadron
are believed to be related to partial restoration of chiral symmetry
in the contemporary point of view
(see, e.g.\cite{review}), in which a reduction of the order
parameter of the chiral phase transition in hot and/or dense matter takes 
place and causes modifications of the hadron properties.
%as excitations on the QCD vacuum.  
The  $N^*$, which is the
lowest lying parity partner of the nucleon, has been investigated from the
point of view of chiral symmetry \cite{DeTar,CDM}, where the $N$ and $N^*$ form a
multiplet of the chiral group.  In refs.\cite{HP,Kim} a reduction of
the mass difference of the $N$ and $N^*$ in the nuclear medium is found in the
chiral doublet model, while a reduction of the $N^*$ mass is also
found as the quark condensate decreases in the QCD sum rule \cite{JKO}. 
Considering the fact that the $N^*$ mass in free space lies only fifty
MeV above the $ \eta N$ threshold, the medium modification of
the $N^*$ mass will strongly affect the in-medium potential of the
$\eta$ meson through the strong $\eta NN^{*}$ coupling described 
above.

As one of the standard theoretical tools to obtain the structure of the bound
states, we solve the Klein-Gordon equation with the meson-nucleus
interaction, which is expressed in terms of an optical potential and is
evaluated using the chiral doublet model developed in refs.\cite{DeTar, CDM}, 
which embodies the reduction of the $N^*$ mass
associated with the partial restoration of chiral symmetry,
assuming the $N^*$ dominance for the $\eta N$ coupling described
above.  The calculation is done in nuclear matter
with the mean-field approximation, and the local density approximation
is used to apply it to the finite nuclei.  For the observation of the these
bound states, we also consider the missing mass spectroscopy by the
(d,$^3$He) reaction which was proven to be a powerful tool
experimentally.  To evaluate the expected spectra theoretically, we
adopt the Green function method for the unstable bound states
\cite{morimatsu85}.

In section 2 we describe the $\eta$-nucleus optical potential which is
obtained in our framework with the chiral doublet model and the
$N^*$ dominance.  
In section 3 we show the numerical results for the $\eta$ meson 
bound states in the nucleus and the  (d,$^3$He) spectra.
Section 4 is devoted to the summary and conclusion.

\section{Optical potential of $\eta$ with $N^*$ dominance}

The $\eta$-mesic nuclei were
studied by Haider and Liu \cite{haider86} and by Chiang, Oset and Liu
\cite{chiang91} systematically.  There, the $\eta$-nucleus optical
potential was expected to be attractive from the data of the
$\eta$-nucleon scattering length and the existence of the bound states
was predicted theoretically.  As the formation reaction, the use of the
($\pi^+,p$) reaction was proposed and the attempt to find the bound
states in the reaction led to a negative result \cite{chrien88}.  The
($\pi^+,p$) experiment \cite{chrien88} was designed to be sensitive to
the expected narrow states, but was probably not sensitive enough to
see  much broader structures.
%%%%%%%

Before presenting the detail of our model calculation, let us show
the possibility to have a repulsive $\eta$ optical potential in the nucleus due to 
the significant reduction of the mass difference of $N$ and $N^*$.  Considering the
self-energy of the $\eta$ meson at rest in nuclear matter in the
$N^*$ dominance model, in analogy with the $\Delta$-hole model for
the $\pi$-nucleus system, we obtain the $\eta$ optical potential  in the 
nuclear medium in the heavy baryon limit \cite{chiang91} as;
\begin{equation}
   V_\eta(\omega) = {\getaNR^2 \over 2 \mu} 
   {\rho(r) \over \omega + \Nmassmed(\rho)  - \Nstarmassmed(\rho) + i
   \Nstarwidmed(s;\rho)/2} \ , \label{poteta}
\end{equation}
where $\omega$ denotes the $\eta$ energy, and $\mu$ is the reduced mass
of the $\eta$ and  the nucleus and is very close to the $\eta$ mass, $\etamass$,
for heavy nuclei. $\rho(r)$ is the density distribution for nucleons in the finite nucleus,
for which we assume here a Fermi distribution.
The ``effective mass'' of $N$ and $N^*$ in medium,
denoted as $\Nmassmed$, $\Nstarmassmed$, are defined through their propagators
so that ${\rm Re\, } G^{-1}(p^0=m^*,\vec p=0)=0$. 
The in-medium $N^*$ width $\Nstarwidmed$ includes the many-body
decay channels.  The $\eta NN^*$ vertex $\getaNR$ is given by
\begin{equation}
   {\cal L}_{\eta NN^*}(x) = \getaNR \bar{N}(x) \eta(x) N^*(x) + {\rm h.c.},
\end{equation}
where $\getaNR \simeq 2.0$ to reproduce the partial width $\Gamma_{N^*
\rightarrow \eta N} \simeq 75$ MeV \cite{PDG} at tree level.

Supposing no medium modifications for the masses  of  $N$ and $N^*$ as well as 
small binding energy for the $\eta$, {\it i.e.} $\omega \simeq \etamass$,
we obtain an attractive potential independent of density since we always
have $\omega + \Nmassvac - \Nstarmassvac < 0$ in the nucleus. The shape
of this potential is essentially the same as the Woods-Saxon type potential for 
a finite nucleus assuming the local density approximation.  On the
other hand, if a sufficient reduction of the mass difference of $N$
and $N^*$ stems from the medium effects, there exists a critical density
$\rho_c$ where $\omega + \Nmassmed - \Nstarmassmed = 0$, and then at
densities above $\rho_c$ the $\eta$ optical potential turns to be
repulsive.  If $\rho_c$ is lower than the nuclear saturation
density $\rho_0$, the optical potential for the $\eta$ is  attractive
around the surface of the nucleus and repulsive in the interior. 
Therefore the bound states of the $\eta$ in nuclei, if they exist, will be
localized at the surface and the level structure of the $\eta$ bound states
will be quite different from that in the case of the
Woods-Saxon type potential.

%%%%%%%%%%%%%%%%%%%%%%%%%%%%%%%%%%%%%%%%%%%%%%%%%

To make the argument more quantitative, we estimate the in-medium $N$
and $N^*$ masses and the $N^*$ width in the chiral doublet model
\cite{DeTar, CDM} (see also \cite{seealso}), which embodies the reduction of the mass
difference of $N$ and $N^*$ associated with the partial restoration of
chiral symmetry.  The chiral doublet model is an extension of the
$SU(2)$ linear sigma model for the nucleon incorporating the $N^*$.  In
the model $N$ and $N^*$ are expressed as  superpositions of $N_1$ and
$N_2$ which are eigenvectors under the chiral transformation.  There
are two types of the linear sigma model choosing either  the
assignment of the same sign of the axial charge to $N_1$ and $N_2$ or
the opposite sign to each other.  The model with the later construction
was first investigated by DeTar and Kunihiro \cite{DeTar}, and was
named the ``mirror assignment'' later on to distinguish it from the first
assignment (``naive assignment'') \cite{CDM}.  Here we shall discuss
only the chiral doublet model with the mirror assignment.  We have
already checked that the naive assignment produces a qualitatively
similar optical potential of the $\eta$ to that with the mirror assignment. 
The quantitative studies with the naive
assignment will be discussed elsewhere.

The Lagrangian of the chiral doublet model with the mirror assignment
is given by 
\begin{eqnarray}
  {\cal L} &=& \sum_{j=1,2} \left[ \bar{N}_j i \delslash N_j -
  g_j \bar{N}_j ( \sigma + (-)^{j-1}i\gamma_5 \vec \tau \cdot \vec \pi) N_j
  \right] \nonumber \\
  && - m_0 (\bar N_1 \gamma_5 N_2 - \bar N_2 \gamma_5 N_1) \ ,
  \label{mirmodel}
  %+ {\cal L}_{\rm meson}
\end{eqnarray}
where $N_1$ and $N_2$ are eigenvectors under the $SU(2)$
chiral transformation and have opposite axial charge to each other.
The physical $N$ and $N^*$ are expressed as  a
superposition of $N_1$ and $N_2$ as $N=\cos\theta N_1 + \gamma_5 \sin
\theta N_2$ and $N^*=-\gamma_5 \sin\theta N_1 + \cos\theta N_2$ where $\tan
2\theta = 2 (g_1+g_2) m_0 /\mevsig $. 
In this basis, the $N$ and $N^*$ masses and the $\pi NN^*$ vertex are
given as functions of the sigma condensate as in the standard liner sigma
model: 
\begin{eqnarray}
    m^{*}_{ N,N^*}
    &=& {1 \over 2} ( \sqrt{ (g_1 + g_2)^2 \mevsig^2 +
    4 m_0^2 } \mp (g_2-g_1) \mevsig ) \label{mass}  \ , \\
    \gpiNR &=& (g_2-g_1)/\sqrt{4 + ((g_1 + g_2)\mevsig/m_0)^2} \ ,
    \label{coup}
\end{eqnarray}
where $\mevsig$ is the sigma condensate in nuclear matter.  It is
worth noting that the mass splitting between $N$ and $N^*$ is
generated by the spontaneous breakdown of chiral symmetry with a
linear form of the sigma condensate.  The parameters in the Lagrangian
have been chosen so that the observables in vacuum, $\Nmassvac=940$
MeV, $\Nstarmassvac=1535$ MeV and $\Gamma_{N^* \rightarrow\pi N}
\simeq 75$ MeV are reproduced with $\langle \sigma \rangle_0 =
f_\pi=93$ MeV, and they are $g_1= 9.8$, $g_2= 16.2$, $m_0 =
270$ MeV\cite{CDM}.

In the linear $\sigma$  model, the nuclear medium effects
come from the nucleon loops which modify the
self-energies of $N$ and $N^*$ and the vertex of $\pi NN^*$ \cite{HKS}.
In the mean-field approximation, such contributions can be introduced by making
the replacement of the vacuum condensate of the sigma meson to the
in-medium condensate which depends on the density $\rho$ as
\begin{equation}
  \mevsig = \Phi(\rho) \langle \sigma \rangle_0 \label{mevsig} \ ,
\end{equation}
where $\Phi(\rho)$ should be determined elsewhere. Here we take
a linear parameterization as $\Phi(\rho)=1-C\rho/\rho_0$ with
$C$=$0.1$-$0.3$ \cite{HKS}. The $C$ parameter represents the strength of 
the chiral restoration at 
the nuclear saturation density $\rho_0$. 
Finally, with this
replacement, the density dependence of the 
in-medium mass difference of $N$ and $N^*$ is obtained as
\begin{equation}
   \Nmassmed (\rho) -\Nstarmassmed (\rho)= (1-C {\rho/\rho_0})(\Nmassvac -
   \Nstarmassvac) \label{massdiff} \ .
\end{equation}
This implies that for $C>0.08$ the real part of the in-medium $\eta$
self-energy with $\omega=m_\eta$ in eq.(\ref{poteta}) changes 
its sign at smaller densities than the saturation density $\rho_{0}$.

Now let us consider the $N^*$ width in the medium. 
In free space there are three strong decay modes of $N^*$, 
$\pi N$, $\eta N$ and $\pi\pi N$ \cite{PDG}. The first two
are dominant modes and give  almost the same decay rate.
Here the medium effects on the decay widths of these modes are 
taken into account by considering the Pauli blocking effect on 
the decaying nucleon, and by changing the $N$ and $N^*$ 
masses and the $\pi NN^*$ coupling in medium
through eqs.\ (\ref{mass}) and (\ref{coup}),
according to the chiral doublet model as discussed above.

The partial decay widths of $N^*$ to certain  decay channels can be obtained
by evaluating the self-energy of $N^*$ associated with the decay
channel and by taking its imaginary part as $\Gamma_{N^*} = -2 {\rm Im}
\Sigma_{N^*}$. 
The partial decay width for $N^* \rightarrow \pi N$ is calculated as
\begin{equation}
  \Gamma_{\pi}(s)=3 {\gpiNRsq \over 4\pi} {E_N + \Nmassmed \over
  \sqrt{s}} q \label{Nstarwid} \ ,
\end{equation}
where $E_N$ and $q$ are  the energy and the momentum of the final
nucleon on the mass shell in the $N^*$ rest frame, respectively. The momentum
$q$ is given by $q=\lambda^{1/2}(s, m_{N}^{*2}, \pimass^2)
/2\sqrt{s}$ with the K\"allen function $\lambda(x,y,z)$. 
The Pauli blocking effect for
$N^* \rightarrow \pi N$ mode is negligible due to the large decaying
momentum $q\simeq 460$ MeV in vacuum.
The decay mode $N^* \rightarrow \eta N$, however, does not
contribute in the medium because of no
available phase space due to the Pauli blocking \cite{chiang91}.

The decay branching rate of the $N^{*} \rightarrow \pi\pi N$ channel is 
known to be only $1 \sim 10$ percent in vacuum \cite{PDG}.
Here in the present calculation, we do not include 
this decay mode since its contribution is estimated to be only a few 
percent of the total decay rate in our model. In spite of 
no direct $\pi\pi NN^*$ vertex in our model, this channel can 
be evaluated by considering the process $N^*
\rightarrow \sigma N \rightarrow \pi\pi N$ using the linear sigma model
for $\sigma$ and $\pi$.

Other medium effects on the decay of $N^*$ are the many-body decays, 
such as $N^*N \rightarrow NN$ and $N^*N \rightarrow \pi NN$. 
The $N^*$ many-body absorption, involving two-nucleon $\eta$
absorption mechanisms, is evaluated by inserting  particle-hole
excitations to the $\pi$ ($\eta$) propagator in the $\pi N$ ($\eta N$)
self-energy of $N^*$.  The width from $NN^* \rightarrow NN$ channel
was already calculated in ref.\cite{chiang91} and was found several
MeV at $\rho=\rho_0$.  Here we neglect the contribution of this 
channel.  The other $N^*$ absorption, $NN^* \rightarrow \pi NN$,
is comparably larger than $NN^* \rightarrow NN$ because of its phase
space \cite{chiang91}.  We estimated the $NN^* \rightarrow \pi NN$
process within our model following the formulation of
ref.\cite{chiang91}:
\begin{eqnarray}
  \lefteqn{\Gamma_{N^*N\rightarrow \pi NN} (\sqrt s)=}    \\ \nonumber 
  &&  3 \beta^2 \left( {g_{\pi NN} \over 2
   m_N^*} \right)^2 \rho \int dp_1 p_1^3 \int {dp_2 \over (2 \pi)^3} 
   p_2  {m_N^* \over \omega_2}  \nonumber \\
  &&  \times
  {-\vec p_1^{\, 2} + 2m_N^* (\sqrt s - \omega_2 - m_N^*) \over 
   \left[ \left( {p_1^2 \over 2 m_N^*} \right)^2 - p_1^2 - m_\pi^2 \right]^2}
  \Phi(p_1,p_2) \ , \label{NNpidecay}
\end{eqnarray} 
where $p_1$ ($\omega_1$) and $p_2$ ($\omega_2$) 
are pion momenta (energies), $\Phi$ is the phase space variable defined in \cite{chiang91} and 
\begin{equation}
\beta = { g_1  m_0 \over \langle \sigma \rangle m_N^* (m_{N^*}^* + m_N^*) }
   \chi 
\end{equation}
with the effective coupling of $ \pi\pi N$ through $\sigma$ meson 
in this model, which is $\chi \sim 1.29$.
This contribution is estimated to be typically fifteen MeV at the
saturation density, although it depends on the $\eta$ energy and $C$
parameter.  We include this channel in the present calculation.

%%%%%%%%%%%%%%%%%%%%%%%%%%
\section{Numerical Results}
%%%%%%%%%%%%%%%%%%%%%%%%%%
In the previous section, we discuss the $\eta$ optical potential in nuclear 
matter using the $N^*$ dominance model and the chiral doublet model.
The final expression of the $\eta$ optical potential in the nuclear medium is
obtained by substituting  the mass gap between $N$ and $N^{*}$ 
obtained in eq.(\ref{massdiff}) and the in-medium $N^*$ width 
described in eqs.(\ref{Nstarwid}) and (\ref{NNpidecay}) to eq.(\ref{poteta}).
In this section, we show some numerical results for the $\eta$ meson 
in the nucleus from the optical potential derived above.

\subsection{Optical potential of the $\eta$ in the nucleus}

\begin{figure}[hbt]
%Fig.1
\epsfxsize=8.5cm
\epsfbox{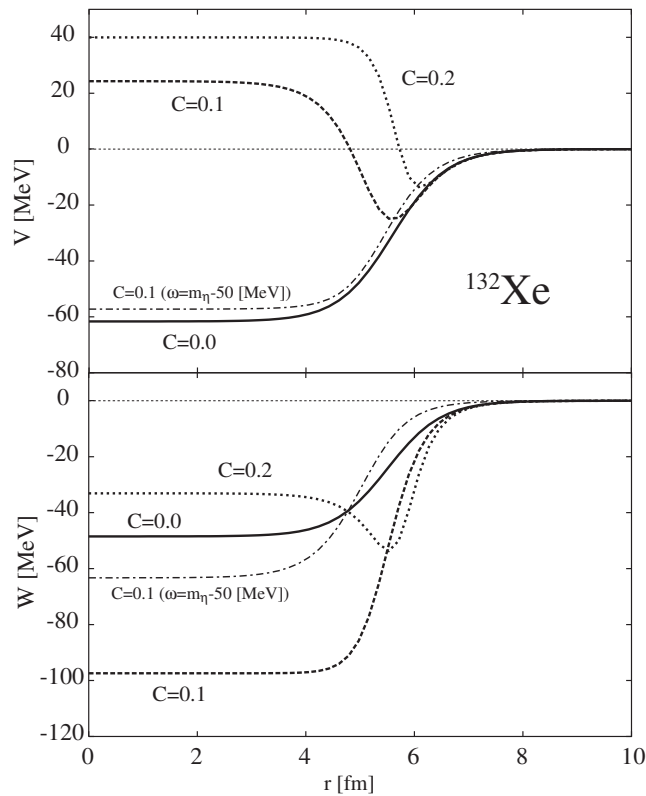}
\caption{The $\eta$-nucleus optical potential for $^{132}$Xe system as
a function of the radius coordinate $r$.  
The upper and lower panels show the real part and
imaginary part, respectively, for $C=0.0$ (solid line), $0.1$ (dashed
line) and $0.2$ (dotted line) with setting $\omega=m_{\eta}$.  The
dot-dashed line indicates the potential strength for $C=0.1$ with
$\omega=m_{\eta} - 50$ [MeV].
\label{potentialfig}}
\end{figure}

For finite nuclei we assume the local density approximation and the Fermi
distribution of nucleons in the nucleus with the radial parameter $R=1.18
A^{1/3}-0.48$ [fm] and the diffuseness parameter $a=0.5$ [fm].  In
Fig.\ref{potentialfig}, we show the $\eta$-nucleus potential for the 
$^{132}$Xe case, as an example.  In other nuclei, the potential shape is 
essentially the same as the plotted one, but the radius of the repulsive core 
depends on the mass number $A$. 
As can be seen in the figure, for the $C\neq 0$ cases
the real potential turns out to be repulsive at the inner part of the
nucleus, associated with the reduction of the mass difference of $N$
and $N^*$ in the nucleus and an attractive ``pocket'' appears on the
surface.  This pocket-shape potential is new and so interesting that the
existence of the repulsive core is consistent with the experiment
\cite{mainz}, where the production of the $\eta$ meson on various nuclei
is surface dominated due to the strong final state interaction. 

Another interesting feature of the potential is its strong energy
dependence.  By changing the energy of the $\eta$ from $\omega =
m_{\eta}$ to $m_{\eta}- 50$ [MeV], we find again the familiar attractive
potential shape even with $C=0.1$ as shown in the figure.  We also find
that the imaginary part of the potential has a strong dependence both on
the $C$ parameter and the $\eta$ energy.

%%%%%%%%%%%%%%%%%%%%%% Figure 1 %%%%%%%%%%%%%%%%%%%%%%%%%

\subsection{Bound states of $\eta$ in nuclei}

In order to calculate the eigenenergies and wavefunctions of the
$\eta$ bound states, we solve numerically the Klein-Gordon equation
with the $\eta$-nucleus optical potential obtained here, and make an
iteration to obtain the self-consistent energy eigenvalues for the
strongly energy dependent  optical potential.
We follow the
method of Kwon and Tabakin to solve it in momentum space
\cite{kwon78}. We increase the number of the mesh points in the
momentum space which is here about 10 times larger than ref.\cite{kwon78}.
The number of mesh points was limited ($\sim$ 40 points) and the
parameters for the mesh points distribution were adjusted for the shallow
mesic atoms \cite{kwon78}.  We check the stability of the obtained results
carefully.  

We show the calculated binding energies and level widths in Table
\ref{eigenenergies} for 1s and 2p states in several nuclei over the
periodic table.  For the $C=0$ case, in which no medium modifications are
included in the $N$ and $N^{*}$ properties, since the potential is
proportional to the nuclear density distribution as we have seen in 
Fig.\ref{potentialfig}, the level structure of the bound states is similar
to that obtained in ref.\cite{hayano99}.  For $C \geq 0.1$ cases, the
formation of $\eta$ bound states is quite difficult 
because of both the repulsive nature of the potential inside the nucleus and 
the huge imaginary part of the potential.

In order to see the $C$ parameter dependence of
the bound states in $C<0.1$, we consider $^{132}$Xe as an example
and calculate the bound states
for $C=0.02$, $0.05$ and $0.08$ cases. 
The results are also compiled in the Table
\ref{eigenenergies}.
We should mention here that, due to the small attraction, in the pocket like case
we do not find bound states. They, however, are obtained in the case of the
Woods-Saxon type.  
% We should mention here that shape of the 
% potentials which give bound states here 
% is not pocket-like but still the Woods-Saxon type due to the
% smallness of the medium effects on $N^{*}$. 

%%%%%%%%%%%%%%%%%%%%%%% TABLE 1 %%%%%%%%%%%%%%%%%%%%%
\begin{table}[ht]
%Table 1
\begin{tabular}{|c|c||c|c||c|c|}
\hline
C&A & \multicolumn{2}{|c||}{ L=0 } & \multicolumn{2}{|c|}{ L=1 } \\
\cline{3-6}
&    &B.E.[MeV]&width[MeV]&B.E.[MeV]&width[MeV] \\
\hline\hline
&6  & 3.7  &  35.1  &    -    &    -     \\ \cline{2-6}
&11 & 13.7  &  41.5  &    -    &    -     \\ \cline{2-6}
&15 & 18.5 &  42.7  &    -    &    -     \\ \cline{2-6}
&19 & 21.9 &  43.1  &    -    &    -     \\ \cline{2-6}
0.0&31 & 27.9 &  42.8  & 10.1 &  52.2  \\ \cline{2-6}
&39 & 30.3 &  42.5  & 14.6 &  50.7  \\ \cline{2-6}
&64 & 34.4 &  41.3  & 22.4 &  47.7  \\ \cline{2-6}
&88 & 36.4 &  40.5  & 26.4 &  45.8  \\ \cline{2-6}
&132& 38.4 &  39.6  & 30.5 &  43.8  \\ \cline{2-6}
&207& 39.9 &  38.5  & 33.9 &  41.8  \\ \hline\hline

0.02&  132 & 41.2 & 49.0  & 33.0 & 55.0  \\ \hline
0.05& 132 & 45.1  & 69.3  & 35.5 & 81.5  \\ \hline
0.08& 132 & 46.1 & 106.3 & - & - \\ \hline
0.1& 132 & - & - & - & - \\ \hline
\end{tabular}
\caption{
The $\eta$-nucleus binding energies and widths for various
nuclei for $C=0$.
Results for the $C\ne 0$ cases 
are also shown for the $\eta$-$^{132}$Xe system. \label{eigenenergies}}
\end{table}
%%%%%%%%%%%%%%%%%%%%%%%%%%%%%%%%%

\subsection{Spectra of (d,$^3$He) for the $\eta$ mesic nuclei formation}
Although the formation of the bound states of the $\eta$ in nuclei is difficult with 
$C\sim 0.2$, which is the expected strength of the chiral restoration in the nucleus,
it would be interesting to see if the repulsive nature of the optical potential can be
observed in experiment. Here we consider the recoilless (d,$^3$He) reaction, in which
a proton is picked up from the target nucleus and the $\eta$ meson is left with 
a small momentum. 

The recoilless (d,$^3$He) reaction is expected to be one of the most powerful
experimental tools for formation of the $\eta$-mesic
nucleus \cite{yamazaki96}.  The spectra of this reaction are
investigated in detail in ref.\cite{hayano99}.  There they estimate
the experimental elementary cross section of $ d + p \rightarrow
^3$He$ + \eta $ reaction to be 150 (nb/sr) in this
kinematics based on the data taken at SATURNE \cite{berthet85}, and
use the Green function method \cite{morimatsu85}, in which the
$\eta$-meson Green function provides information of the structure of
unstable bound states.  Applying the same approach, we evaluate the
expected spectra of the (d,$^3$He) reaction by assuming our optical
potential for the $\eta$ in the nucleus.
Here we calculate the spectra of the $^{12}$C(d,$^3$He) reaction for the 
$\eta$ production in the final state. The $^{12}$C is shown
to be a suitable  target to populate the [$(p_{3/2})^{-1}_p \otimes
p_{\eta}$] configuration largely \cite{hayano99}. 
We assume the single particle nature for the
protons in the target $^{12}$C and consider $0s_{1/2}$ and $0p_{3/2}$ 
states. And we take into account sufficient number of partial waves $l_\eta$ of 
$\eta$. We find that only $l_\eta \leq 3$ partial waves are relevant in this 
energy region and we check numerically that contributions from 
higher partial waves are negligible. 

\begin{figure}[htb]
%Fig. 2
\begin{center}
\epsfxsize=8.5cm
\epsfbox{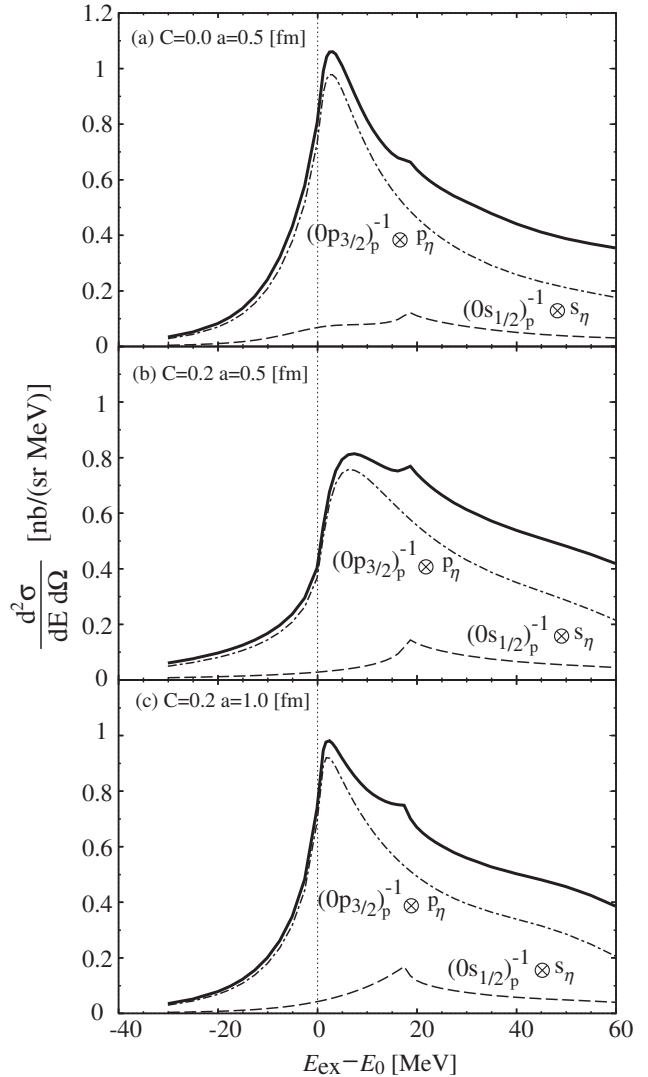}
\caption{
The calculated excitation energy spectrum of the $\eta$ production
in the $^{12}$C(d,$^3$He) reaction at $T_d=3.5$ [GeV] for three different
sets of 
the parameter $C$ in the $\eta$-nucleus optical potential 
% the $\eta$-nucleus optical potential parameter $C$ 
and the diffuseness parameter $a$ of the nuclear density;
(a) $C=0.0$ and $a=0.5$ [fm], (b) $C=0.2$ and $a=0.5$ [fm], (c) $C=0.2$ 
and $a=1.0$ [fm].
The vertical line  indicates the $\eta$ production threshold energy.
In each figure, the contribution from the $(0p_{3/2})_p^{-1} \otimes
p_{\eta}$ is shown in a
dash-dotted curve, the $(0s_{1/2})_p^{-1} \otimes s_{\eta}$ contribution is
shown in
the dashed curve, and the solid curve is the sum of $\eta$-partial waves.  The
continuum
background contributions are estimated to be about 3.4 [nb/(sr MeV)] for the
$^{12}$C target
\cite{hayano99}.
\label{fig:spec}}
\end{center}
\end{figure}

The obtained spectra are 
shown in Fig.\ref{fig:spec} as functions of the excited energy which are
defined as,
\begin{equation}
    E_{\rm ex} = m_{\eta} - B_{\eta} + (S_{p}(j_{p}) - S_{p}(p_{3/2})) \ ,
\end{equation}
where $B_{\eta}$ is the $\eta$ binding energy and $S_{p}$  the
proton separation energy.  The $\eta$ production threshold energy
$E_{0}$ is indicated in the figure by the vertical line.  
The calculated spectra are shown in Fig.\ref{fig:spec} for three different sets of 
the parameter $C$ in the optical potential 
and the diffuseness parameter $a$ of the nuclear density
distribution.
In Fig.\ref{fig:spec}(a), we show the result with $C=0.0$ and $a=0.5$ [fm] which
corresponds to the
spectrum with the potential which does not include any
medium modifications of $N$ and $N^*$.
The results with the medium corrections are shown in Fig.\ref{fig:spec}(b) 
for the $C=0.2$ case, where
the $\eta$ optical potential has the repulsive core in the center of nucleus.
It is seen in the Fig.\ref{fig:spec}(a) and (b) that, due to the repulsive nature 
of the $\eta$ potential with $C=0.2$, the whole spectrum is shifted to the 
higher energy region
and the $s$-wave $\eta$ contribution around $E_{\rm ex}-E_0 \sim$ 0 [MeV] is
suppressed in Fig.\ref{fig:spec}(b), which corresponds to the disappearance of the 
$\eta$-bound state for $C=0.2$.
The difference of these spectra is expected to be observed 
in the high resolution experiments. 

We also calculate the spectrum for 
the $C=0.2$ case  with the diffuseness $a = 1.0$ [fm], 
to simulate the halo-like structure of unstable nuclei.  In this case,
the $\eta$-nucleus optical potential has a wider attractive region than
that with $a=0.5$ [fm] because of the existence of the longer tail at 
low nuclear densities.  The results are shown in Fig.\ref{fig:spec}(c), 
where we can see 
that the whole spectrum is shifted to smaller energy regions compared to 
Fig.\ref{fig:spec}(b) indicating the attractive nature of the potential.

%%%%%%%%%%%%%%%%%%%%%% Figure 2 %%%%%%%%%%%%%%%%%%%%%%%%%

\subsection{Discussion}
Here we make some remarks on the ''pocket-like" potential. 
First of all, we would like to consider the interesting
possibility to have $\eta$ bound states in $\beta$-unstable nuclei
with a halo structure \cite{tanihata96}. 
So far, we have considered stable nuclei with the small
diffuseness parameters ($a=0.5$ [fm]), in which the density suddenly
changes at the surface. As we have seen in Fig.\ref{potentialfig}, 
the optical potential with a finite $C$ value is
attractive only in the low density region at the nuclear surface.  Thus, halo
nuclei have a larger attractive region than the stable nuclei and we
expect to find more $\eta$ bound states, which will be like
coexistence states of the halo-like $\eta$ mesic bound states with
halo nucleons.  We simulate the behavior of the $\eta$ bound states
for nuclei with halo structure by changing the diffuseness parameters
artificially for $^{132}$Xe case.  However, in spite of our systematic
search of the $\eta$ bound states for diffuseness parameters, $a=1.0$ and
$1.5$ [fm], no bound states in the pocket-like potential are found for the 
$C=0.1, 0.15$ and $0.2$ cases.

Second, we have another quite interesting feature of the
$\eta$-nucleus bound states because of its strong energy dependence. 
In the iterative calculation to get the self-consistent eigenenergies
for the energy dependent potential, we have possibilities to find
several solutions for certain set of the quantum numbers $(n,l)$. 
Actually, we have found the three 1s-like states for the $\eta$-$^{88}$Y
system for the $C=0.1$ case by omitting the imaginary part of the optical
potential.  The wavefunctions of all three states do not have any node
and they are shown to be; (1) The bound 1s state with ${\rm B.E}.=40.4$ [MeV]
in the Woods-Saxon type potential, (2) The bound 1s state with ${\rm B.E.} =22.2$
[MeV] in the attractive potential which is deeper at the surface than in the nuclear
center, (3) The bound 1s state with ${\rm B.E.}=6.0$  [MeV] in the
surface-attractive and central-repulsive potential.

Third, we would like to make a general comment on the level structure of the 
bound states in the pocket-like potential.  
To make the argument clear, we consider only the real 
part of the pocket-like potential.
Let us consider the radial part of the Schr\"odinger equation with an
attractive pocket potential at a certain radius $R$ and a width
$a$.
As we can see in any text book of quantum mechanics, this equation is
exactly the same as the one dimensional Schr\"odinger equation with the
same potential pocket except for the centrifugal potential and the
boundary condition for the wavefunction at the origin $r=0$.  Since
the one dimensional Schr\"odinger equation has translational
invariance with this potential, the eigenenergies do not depend on
the position of the attractive pocket.  This is also the case for the
radial equation for heavy nuclei in which the attractive potential exists
far from the nuclear center, where the wavefunction boundary condition
is automatically satisfied and do not affect the eigenenergies.  Thus,
the level structure would be expected to  resemble each other for
all heavy nuclei, which is quite different from the case of the 
Woods-Saxon type potential, where the binding energies become larger in 
heavier nuclei.  

Another interesting feature of the level structure  is that, since 
the pocket potential in
heavy nuclei is far from the center of the system and then the
centrifugal repulsive potential could be weak at the position of the
pocket, then the different angular momentum states would approximately
degenerate in heavy nuclei.  All these features of the level structure are very
interesting and really characteristic for the 'surface-attractive'
$\eta$-nucleus optical potential.

Very recently an investigation of the $\eta$ meson properties
in the nuclear medium within a chiral unitary approach has been reported \cite{inoue}. 
There, they also found a strong energy dependent optical potential of the 
$\eta$.  It would be  interesting to compare their consequences with ours
since their theoretical framework is quite different from our model
and there the $N^{*}$ is introduced as a resonance generated
dynamically from meson-baryon scatterings.

\section{Summary}

We investigate the consequences of the medium effects to $N(1535)$
$(N^*)$ through the eta-mesic nuclei assuming the $N^{*}$ dominance in the
$\eta$-$N$ system.  The chiral doublet model is used to estimate the
medium modification of the $N$ and $N^{*}$ properties.  This model
embodies chiral symmetry and its spontaneous breaking within the
baryonic level ($N$ and $N^{*}$) and shows the reduction of the mass
difference of $N$ and $N^{*}$ with the partial restoration of chiral
symmetry.

We find that sufficient reduction of the mass difference due to chiral
restoration makes the $\eta$ optical potential in nuclei 
repulsive at certain densities, while in the low density approximation
the optical potential is estimated to be attractive.  This leads us the
possibility of a new type of potential of the $\eta$ in nucleus
that is attractive at the surface and has a repulsive core at the center
of the nucleus.  We discuss general features of this ``pocket'' potential
that the level structure is quite different from that with the
potential simply obtained from the $V \sim t \rho$ approximation with $t$
the scattering amplitude  of the $\eta$-$N$ system.

We calculate the in-medium optical potential of the $\eta$ meson with
a mean-field approximation and use the local density approximation to
apply it to finite nuclei.  We also find that the potential obtained here
has a strong energy dependence, which leads us to the self-consistent
formulation in the calculation of the $\eta$ bound states  in the 
nucleus.  Unfortunately it is hard to form $\eta$ bound states in the
nuclei with the expected strength of the chiral restoration in
nucleus ($C \sim 0.2$), due to the repulsive nature of the potential inside the nucleus
and its large imaginary potential.
We find that there is no bound state for
the case of $C\ge 0.1$ even in the halo-like nuclei, which are expected
to have moderate density distributions and to have a wider attractive
potential pocket at the surface.

We evaluate the spectra of the recoilless (d,$^{3}$He) reaction with a
$^{12}$C target using our optical potential for the $C=0.0$ and $0.2$ 
cases.  The shapes of these spectra are apparently different and the 
repulsive nature in the $C=0.2$ case is seen.  
The results with the large diffuseness parameter for the nuclear density
distribution also show  the sensitivity of the spectrum to the
existence of the halo structure.  We find that the existence of the halo
also changes the spectrum shape, and thus a comprehensive study of
$\eta$ bound states in unstable nuclei also would be interesting to
investigate the medium effects of the $N^*$.
We believe that the present results are very important to investigate the
chiral nature of $N$ and $N^*$ through $\eta$ bound states.

% \begin{figure}
% \epsfbox{fig1.eps}
% \caption{}
% \end{figure}

%\section*{Acknowledgment}
\begin{acknowledgments}
We would like to thank Prof. E. Oset for valuable discussions and carefule 
reading of the manuscript. 
The work by D.J. was supported by the
Spanish Ministry of Education in the Program ``Estancias de Doctores y
Tecn\'ologos Extranjeros en Espa\~{n}a''.
One of us, S.H. wishes to acknowledge the hospitality of the University
of Valencia where this work was done and financial support from the Foundation
BBV. This work is also
partly supported
by the Grants-in-Aid for Scientific Research of the Japan Ministry of
Education, Culture, Sports, Science and  Technology (No. 14540268 and No.
11694082).
\end{acknowledgments}

%%%%%%%%%%%%%%%%%%%%%%%%%%%%%%%%%%%%%%%%%%%%%%%%%%%%%%%%%

\end{document}